\documentclass[lettersize,journal]{article}
\usepackage{amsmath,amsfonts}
\usepackage{algorithmic}
\usepackage{algorithm}
\usepackage{array}
\usepackage[caption=false,font=normalsize,labelfont=sf,textfont=sf]{subfig}
\usepackage{textcomp}
\usepackage{stfloats}
\usepackage{url}
\usepackage{verbatim}
\usepackage{graphicx}
\usepackage{cite}
\usepackage{xcolor}
\usepackage{mathtools}

\providecommand{\keywords}[1]
{
  \small	
  \textbf{\textit{Keywords---}} #1
}

\newtheorem{remark}{Remark}

\begin{document}

\title{Autonomous Emergency Braking With Driver-In-The-Loop: Torque Vectoring for Active Learning }

\author{Benjamin Sullivan, Jingjing Jiang,  Georgios Mavros, Wen-Hua Chen
        % <-this % stops a space
\thanks{
% Manuscript received April 19, 2021; revised August 16, 2021.
The authors are with the Department of Aeronautical and Automotive Engineering, Loughborough University, LE11 3TU, UK. Correspondence; b.sullivan@lboro.ac.uk; j.jiang2@lboro.ac.uk;  g.mavros@lboro.ac.uk; w.chen@lboro.ac.uk.
This work has been submitted to the IEEE for possible publication. Copyright maybe transferred without notice, after which this version may no longer be accessible.
This work was supported by the UK Engineering and Physical Sciences Research Council (EPSRC) Established Career Fellowship ‘‘Goal-Oriented Control Systems: Disturbance, Uncertainty and Constraints’’ under the grant number EP/T005734/1.
}}



\date{}
\maketitle

\begin{abstract}
Autonomous Emergency Braking (AEB)  brings significant improvements in automotive safety due to its ability to autonomously prevent collisions in situations where the driver may not be able to do so. 
Driven by the poor performance of the state of the art in recent testing, this work provides an online solution to identify critical parameters such as the current and maximum friction coefficients.
% parameters of the Magic formula tyre model, required for emergency braking. 
% Furthermore we do this without disturbing the driver and in the presence of measurement noise.
The method introduced here, namely Torque Vectoring for Active Learning (TVAL), can perform state and parameter estimation whilst following the driver's input. 
% Importantly with less power requirements than normal driving.
%Importantly we do not disturb the driver whilst using our method, leaving them to control the vehicle.
Our method is designed with a crucial focus on ensuring minimal disruption to the driver, allowing them to maintain full control of the vehicle.
Additionally, we exploit a rain/light sensor to drive the observer resampling to maintain estimation certainty across prolonged operation.
Then a scheme for TVAL is introduced which considers powertrain efficiency, safety, and feasibility  in an online fashion.
Using a high-fidelity vehicle model and drive cycle, we demonstrate the functionality of the TVAL controller across changing road surfaces, where we successfully identify the road surface whenever possible.

% Driven by the limited research in this area and poor performance of commercial systems we introduce Torque Vectoring for Active Learning (TVAL) to improve AEB.
\end{abstract}


\section{Introduction}
Autonomous Emergency Braking (AEB) systems are gradually penetrating into main stream, mass production vehicles. However literature is limited with significant problems reported in testing of commercial systems. Recent (2022) testing conducted by the American Automobile Association (AAA) and National Highway Traffic Safety Administration (NHTSA) found that, at an approach speed of \(40mph\), AEB systems only prevented \(30\%\) of collisions and in some common scenarios, preventing collisions was not achieved at all \cite{AAA2022}.
To ensure safe vehicle control, real-time understanding of the interaction between the tyres and the road surface is critical. 
Advanced driver assistance features, such as the Anti-Lock Brake system (ABS) \cite{sullivan2023} or Electronic Stability Control (ESC) are required to control the vehicle at the limits of adhesion which can rely on accurate identification of the vehicle states and tyre model when maximizing the tyre friction. 
In the case of more automated or autonomous features, such as AEB, this information must be known prior to the emergency braking and with a high degree of confidence to avoid collision.
Overestimation of the current and maximum achievable friction leads to collision since the controller believes that greater deceleration is possible. 
Under-estimation results in poor confidence and creates a lack of trust from the driver in the vehicles active safety features, potentially leading to misuse or deactivation. 
The optimal identification of the tyre model and road surface is challenging due to the wide range of operating conditions a typical vehicle may experience. 
These include varying road surfaces e.g. from dry to wet to icy, alongside potential changes in  tyre behaviour stemming from factors like  tread wear, temperature fluctuations or inflation variations.
Additionally, the vehicle may operate  across a wide range of states, spanning  high to low velocities.
Our method, namely Torque Vectoring for Active Learning (TVAL) addresses these obstacles and identifies both the current and maximum friction coefficient whilst reducing the uncertainty.
Since it is important to maintain a high level of confidence when taking control of the vehicle during  emergencies, 
% Our method, Torque Vectoring for Active Learning (TVAL) can adapt across changing operating conditions and yield valuable tyre model and vehicle state information, useful for active safety features, such as AEB. 
 our method stands out by conducting the identification process preemptively, before such situation arise. 
 This approach  can  accurately estimate the optimal stopping distance while minimizing any perturbations on the driver.

% \IEEEpubidadjcol

While direct observation of the friction coefficient through tire force sensors, as demonstrated in \cite{Mazzilli2021}, is possible, it proves prohibitively expensive and impractical for mass production vehicles.
Our work establishes that such an approach is  unnecessary, since we achieve comparable results  with low-cost, readily  available sensors. 
Friction estimation is primarily considered by existing research using passive means, that is, using the input from the driver to learn the friction state. 
These use either local (effect) methods to estimate the current friction at the tyre contact patch, or predictive (causal) methods to find the future maximum friction coefficient.
% Although our method is local, it can be used for predictive means since we identify a complete tyre model and could hence use it to predict future friction conditions if the maximum friction were to remain constant.
Predictive methods overwhelmingly utilize cameras \emph{e.g.} \cite{Casselgren2012}, to identify and segment the road surface according to its friction properties.  
It has been acknowledged that optical based methods alone are unable to determine the friction coefficient with sufficient accuracy \cite{Svensson2021}. 
These methods largely rely on classifying road conditions into a limited set of categories with attributed friction properties.
In practice, this approach can lead to substantial estimation errors. 
For example, a wet road may have a maximum friction coefficient between \(0.3-0.9\) \cite{S2015}. In \cite{Cheng2019} the road type classification using 5 categories of road surface is demonstrated to have an accuracy of 95\%. 
\cite{Wang2022} developed an AEB estimation method to jointly estimate the road grade and road type using dry, wet or snow classifiers with between \(75\%-85\%\) identification success. 
% If we perform a simple braking example using this error in our peak friction estimation, assuming dry road surface and \(94\%\) estimation accuracy, this leads to a worst case scenario of \(0.65m\) longer stopping distance (\(6\%\)) if braking from \(40\emph{mph}\), which isn’t acceptable.

Early methods for local friction estimation initially focused on whole vehicle friction estimation, providing an average estimate for the whole vehicle, as seen in works such as \cite{Gustafsson1997} \cite{Yi1999}.
This approach later evolved into  methods that target individual wheel estimation \emph{e.g.} \cite{Rajamani2012}. 
In \cite{Ray1997}, a Kalman Filter was used to directly identify the current tyre friction coefficient but without identifying a tyre model. 
A novel tyre model was developed in \cite{Zhang2022} and was successfully identified using a Square Root Cubature Kalman Filter but requires manufacturers to parameterize their tyres according to the new model. 
Similarly \cite{Djeumou2023} presents an offline method using 3 minutes of drifting making it unsuitable for normal driving.
Others  identified the Brush tyre model which requires fewer parameters but comes at the expense of accuracy \cite{Ma2018}.
In comparison, our work identifies the widely accepted Magic Formula tyre model which can model a large range of tyres. 
% An offline approach is shown in in \cite{Djeumou2023} that requires 3 minutes of aggressive drifting maneuvers to parameterize custom tyre models. 
% Using neural ordinary differential equations, high model parametrization accuracy could be achieved at speeds up to \(45\) \emph{mph} although the nature of the driving behavior makes this impractical for normal road vehicles.

There has been a lot of interest in passive identification methods that use regular driver inputs to identify friction coefficients. However, active methods require additional vehicle inputs such as braking or driving torque to identify coefficients at large slip angles.
This becomes particularly necessary when the relationship becomes nonlinear (where peak friction occurs). 
Active learning through tyre force excitation is known to improve the estimation quality \cite{Prokes2016}. 
However very few researchers acknowledge this by designing active learning friction identification methods. 
The aim of these methods is to perturb the behavior of the wheel above what the driver provides as inputs to the vehicle but without disturbing the driver. 
The authors of \cite{Albinsson2017} acknowledge this and investigate how different excitation strategies impact the accuracy of longitudinal-only vehicular motion identification. 
Although their findings offer useful insights, such as the choice of tyre model required different levels of excitation, with the Magic formula requiring the least excitation, their optimization approach is not suitable for real-time control due to its reliance on the batch approach. 
% In practice, an iterative approach is necessary since the road and tyre are dynamic and hence constantly changing. 
Moreover, their work overlooks the issue of determining when tyre force excitation should begin, causing a potential safety concern in real-world implementation, which our study addresses. %and provides a barrier for physical implementation since it may not be safe to always be actively exciting the vehicle, although is addressed in our work. 
In \cite{Zhao2023}, an attenuated excitation signal is designed to provide sufficient levels of excitation. However, this does not include the level of uncertainty in the estimation; instead, an attenuation index parameter is manually selected for estimation performance.
An approach using PID control is shown in \cite{Albinson2016}, but it only tracks a step torque input which is supplied to each wheel (with opposite signs) and achieves an acceleration error of \(0.7m/s^2\). 
Since opposite torques are used, tyre forces become unbalanced at high values of slip and asymmetric vehicle loading.
An offline nonlinear least squares method was used to fit the measurements to a brush tyre model.
Among the literature surveyed, none of the methods accomplishes this identification whilst considering a non-zero acceleration input from the driver; however, our method successfully achieves this capability.
%Our scheme continuously balances between controls that improve the estimation performance and those that drive the vehicle according to the driver's inputs.
Existing research is limited when considering the human-machine interface \cite{Picotti2023} and thus our further motivates our work.

% It is more useful for vehicle control to have not only the knowledge of the current friction but also the knowledge of the maximum friction, useful vehicle handling at the limits of adhesion, and also knowledge of the friction coefficient at any state of the vehicle. 
Our method is well equipped to work with similar vehicle control systems that utilize torque vectoring, namely those used for stability control, which has increased growth in the last 5 years due to the adaptation of electrified vehicles \cite{Warth2019}. 
These methods encounter challenges as they necessitate highly accurate prior tire and vehicle information for effective operation.
% \textcolor{red}{explain these simple facts along the following lines first:
% \begin{enumerate}
% \item{road friction pays the profound effect in braking and the minimum stopping distance;}
% \item{it is important to estimate the road friction and its associated uncertainty so decide when to activate AEB;}
% \item{many factors influence it including road surface, weather conditions, tyre conditions;}
% \item{it could not be estimated offline so a capability of real-time monitoring road friction is essential for AEB}
% \end{enumerate}
% }

% Previously the Dual Control for Exploration and Exploitation (DCEE) method \cite{sullivan2023} was used in an emergency braking scenario for an Anti-lock Brake System (ABS) where the whole vehicle was driven in a exploration/exploitation manner to actively learn the system. 
% *Different from ABS, AEB introduces the requirement to know the state of the road surface ahead of time and without disturbing the driver. 
% Thus, this work aims to develop a control system that performs the estimation tasks using behaviours that do not impact the drivability of the vehicle but provide an accurate estimation of when to autonomously activate the brake.

\begin{figure}
    \centering
    \includegraphics[scale=0.43]{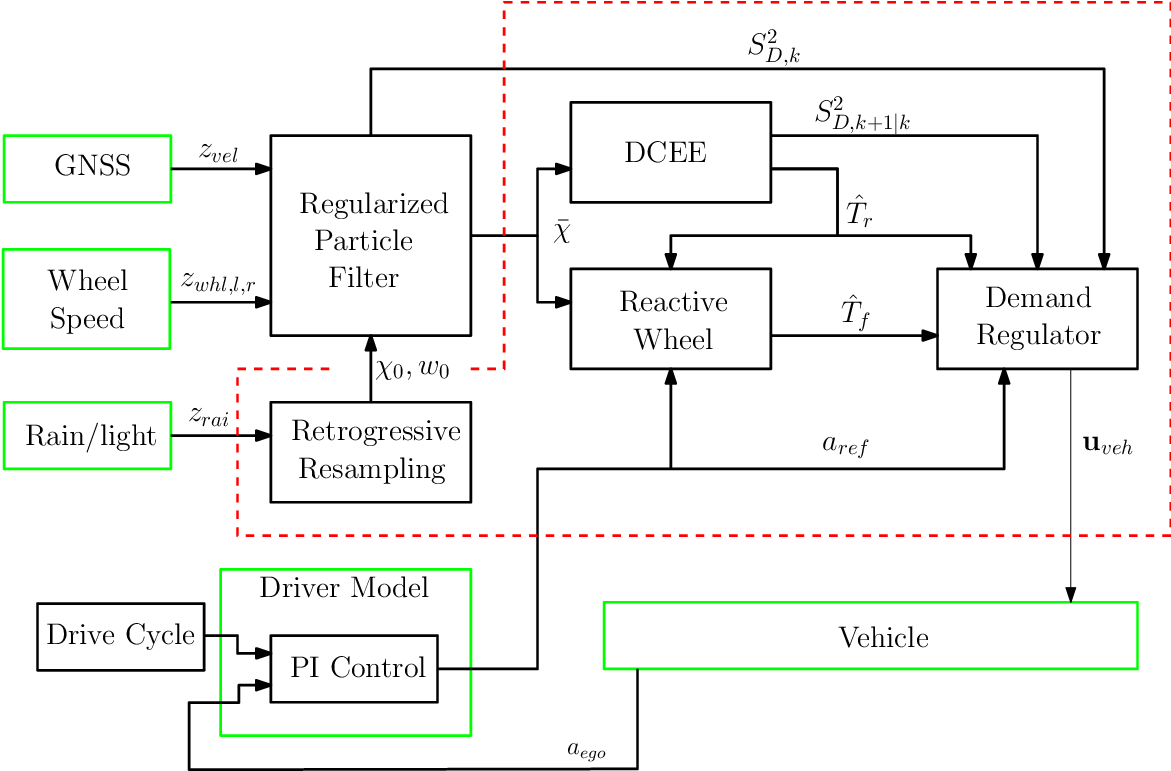}
    \caption{TVAL system architecture for AEB with the novel contribution shown in red. A Global Navigation Satellite System (GNSS) \emph{e.g.} GPS gives vehicle speed measurements and Dual Control for Exploration-Exploitation (DCEE) is used as part of our control strategy.}
    \label{fig:SAD}
\end{figure}

The contributions of this work are as follows.
We present a novel TVAL control scheme (see Fig.~\ref{fig:SAD}) that identifies the current and maximum road surface friction coefficients while following the driver's input. 
This is done by identifying  the parameters of the Magic formula in a gain free manner.
Unique to this work, rain/light sensor (used for windscreen wipers) is used as part of a resampling scheme to drive the on-line estimator.
Then, we introduce a new method for regulating active learning by incorporating considerations of tyre force availability, energy efficiency and safety. 
This enhancement significantly augments the applicability of the scheme for real world deployment. 
Notably, a hysteresis switch-based strategy for intelligently distinguishing when TVAL can improve the system belief of the peak friction coefficient is presented. 
%Then a method for predicting the oversteer gradient of the vehicle when using TVAL as to prevent instability in the event of unforeseen lateral disturbance whilst doing TVAL.
A vehicle dynamics model is also used to predict the vehicle's understeer gradient during TVAL, effectively mitigating instability caused by unexpected lateral disturbances.
A smooth transition between active learning and purely driver inputs is also introduced. 
Finally, the complete control scheme is demonstrated in simulation using a validated 7DOF vehicle dynamics model of a Jaguar XE and a driver model. 
%The human driver is replicated using the Extra-Urban Drive Cycle (EUDC) and a Proportional Integral controller. 
Here a mixture of road surfaces, including dry, wet, and snowy road surfaces, is investigated.
%The control system overview is shown in Fig. \ref{fig:SAD}.

The rest of the paper is organized as follows. In Section \ref{sec:2} the vehicles dynamic and tyre model are presented. Then in Section \ref{sec:3} a motivation for active learning is shown with TVAL being introduced in Section \ref{sec:4}. 
Section \ref{sec:5} presents the full TVAL scheme, with simulation demonstrations provided in Section \ref{sec:6}, and conclusions drawn in Section \ref{sec:7}.

\section{Vehicle Dynamics}\label{sec:2}
\begin{figure}
    \centering
    \includegraphics[scale=0.6]{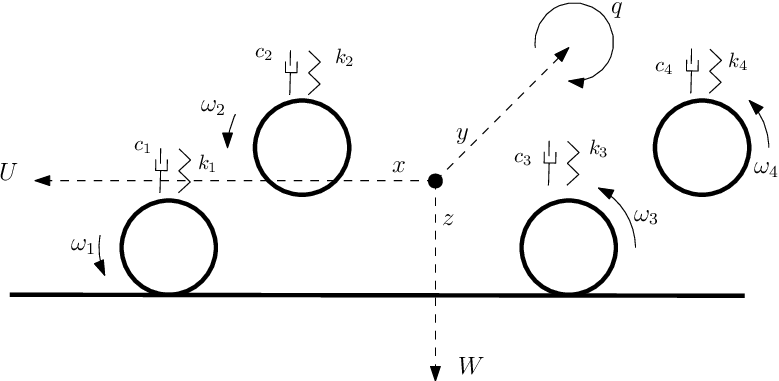}
    \caption{7 DOF  vehicle dynamics model.}
    \label{fig:vehicle}
\end{figure}

This work focuses on the braking only behaviour of the vehicle, hence we focus on the 7 DOF shown in  Fig.~\ref{fig:vehicle}. 
The model used in our analysis follows the SAE (Society of Automotive Engineers) frame of reference \cite{Dixon1996} and was validated against a real vehicle \cite{Mavros2008}.

\subsection{Vehicle body}
The contribution of all longitudinal (\(x\) direction) and vertical (\(z\) direction) tyre forces \(F_{x,j}\) is found using the vehicle mass \(m\) while considering the movement of the centre of gravity (COG), denoted as \(\left[x_G,y_G,z_G\right]\), indicating
\begin{equation}\label{eqn:fx}
\sum^4_{j=1} F_{x,j}=m \left( \frac{dU}{dt} + Wq \right)-m \left(x_G q^2 - z_G \frac{dq}{dt}\right)
\end{equation}
\begin{equation}\label{eqn:fy}
\sum^4_{j=1} F_{z,j} =
m \left( \frac{dW}{dt} - Uq \right) - m \left( z_G q^2 + x_G \frac{dq}{dt}\right)
\end{equation}
where \(U\) and \(W\) are the velocities of the vehicle body (at the COG) in the longitudinal and vertical directions, \(q\) is the rate of change in pitch of the vehicle body and \(j\) denotes the location of the wheel, \emph{i.e}, \(j \in \left[1,2,3,4\right]\).
%Where \(m\) is the vehicle mass. 
Furthermore, the rotation about the y-axis can be calculated by (\ref{eqn:my}) using the moment of inertia about the \(yy\) axis, \(I_{yy}\). The moments produced by the forces at each wheel, \(M_{y,j}\) are found as:
\begin{equation}\label{eqn:my}
\sum^4_{j=1} M_{y,j} = I_{y y} \frac{dq}{dt} + m z_G \left( \frac{dU}{dt} + Wq \right) - m x_G \left( \frac{dW}{dt} - U q \right)
\end{equation}
Next we introduce the suspension model using spring stiffnesses for the front and rear \(k_{f,r}\) and damping coefficient \(c_{f,r}\):
\begin{equation}\label{eqn:sus1}
F_{z,f}=k_{\mathrm{f}}\left(z_G-|a|\phi\right)+c_{\mathrm{f}}\left(W-|a| q\right)
\end{equation}
\begin{equation}\label{eqn:sus2}
F_{z,r}=k_{\mathrm{r}}\left(z_G+|b|\phi\right)+c_{\mathrm{r}}\left(W+|b| q\right)
\end{equation}
where \(a\) and \(b\) are the distances between the front and rear axles to the COG such that the sum is equal to the wheelbase \(l\). Additionally the pitch angle is denoted as \(\phi\).
Although this is a first-order system, the major physical characteristics are captured by this relationship.
In the simulation plant model, anti-squat geometry is included which provides more realistic load transfer.

\subsection{Wheels and Tyre Model}
\begin{figure}
    \centering
    \includegraphics[scale=0.81]{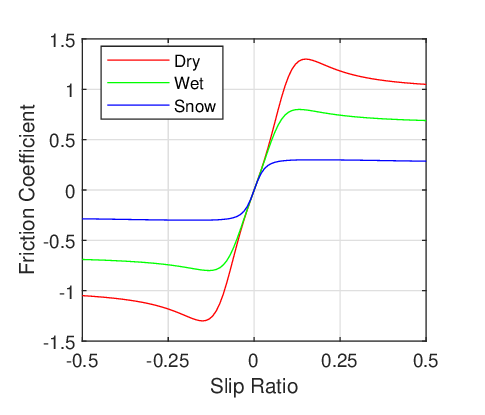}
    \caption{Wheel slip - friction coefficient relationship for positive and negative slips.}
    \label{fig:muslip}
\end{figure}

The wheel dynamics are found using the wheel rotational acceleration, \(\dot{\omega}_j\) and the contribution from braking or accelerating the vehicle, applied as torque \(T_j\).
Hence,
\begin{equation} \label{eqn:wheeldyn}
    I_w\dot{\omega}_j = T_j - RF_{x,j},
\end{equation}
 where \(I_{w,j}\) is the wheel inertia and \(R\) is the radius of the wheel. 
The tyre force is simply the product of the wheel vertical load and the friction coefficient between the tyre and road surface:
\begin{equation}\label{eqn:tforce}
    F_{x,j} = \mu(\kappa_j) F_{z,j}.
\end{equation}
In this work the Magic formula tyre model is used with parameters \(B,C,D\) and \(E\) that are the stiffness, shape, peak\footnote{The peak factor is the maximum available friction on the current road surface} and curvature factors respectively, as 
\begin{equation}\label{eqn:magic}
    \mu_{j}=D\sin[C\arctan({B\kappa_{j}-E(B\kappa_{j}-\arctan B\kappa_{j})})]
\end{equation}
Note that we consider a single tyre model for all four tyres, but each wheel slip is independent from one another. 
%Note the peak factor is the maximum available friction on the current road surface.
Using this tyre model allows for the encompassment of many different tyres on different road surfaces including dry, wet and snow.
Then the longitudinal slip ratio, \(\kappa_j\), is generated as the tyre carcass begins to deform because of the difference in linear speed of rolling and body velocity: %otherwise known as slip:
\begin{equation}\label{eqn:slip}
    \kappa_{j} = \frac{\omega_{j} R - U}{U}
\end{equation}
The relationship between the road-tyre friction coefficient and wheel slip is shown in Fig. \ref{fig:muslip} for both braking and driving cases (\emph{i.e.}, negative and positive slip respectively).

\section{Passive vs Active Learning with Dual Control}\label{sec:3}
An estimator may also be able to quantify the uncertainty of an estimation, \emph{e.g.} covariance matrix in Kalman filter or particle weights in a Particle filter, when evaluating a state or parameter. Yet no explicit action is taken to reduce this so long as the control reference/objective is achieved. 
Therefore,  traditional control methods often fail to address the issue of cognitive dissonance, characterised by  the discrepancy between system belief and observation, when selecting an appropriate control policy. 
Similarly, the separation principle whereby an optimal controller and an optimal observer are designed separately is often utilized \emph{e.g} \cite{Qin2022}. 
%due to non-linear plant dynamics and real-world sensor characteristics. 
However, this introduces more instability in the observer which causes issues in the controller when tracking an estimated state.
Dual Control for Exploration and Exploitation (DCEE) is well placed in this regard since it actively takes actions to reduce the \textit{stress} caused by persistent uncertainty of working in unknown environments whilst following an objective (duality) (\emph{e.g}\cite{Chen2021}). 

%Traditionally, Dual Control \cite{feldbaum1960dual}  was intractable and difficult to achieve real time implementation, while DCEE is not.
%Furthermore, 
In practical goal orientated control systems, the reference may not be known, yet only the high level mission is specified. 
Hence the reference is required to be estimated during the operation of the controller and within an unknown or unfamiliar environment. 
In this sense, the DCEE approach is more beneficial than Reinforcement Learning that is limited when dealing with changing tasks \cite{Chen2022}.
DCEE takes this idea of an initially unknown reference into consideration and due to the duality of the method, the uncertainty associated with it is in an exploration/exploitation manner.
% \textcolor{red}{again DCEE is a rather new concept and you cannot presume the reviewers/reader are fully knowledgeable about it. You need to introduce it and justify why it is suitable for your problem before going to the detail}
Furthermore, TVAL makes improvements on the issue of cognitive dissonance by considering both the estimation uncertainty and prediction uncertainty when deciding on whether to do active learning or not. 
%which active learning policy is appropriate.
% Artificially Intelligent (AI) methods such as Neural Networks have shown useful when working in uncertain environments yet due to their 'black box' nature are challenging to guarantee the performance of.

% Active inference
\begin{figure}
    \centering
    \includegraphics[scale=0.78]{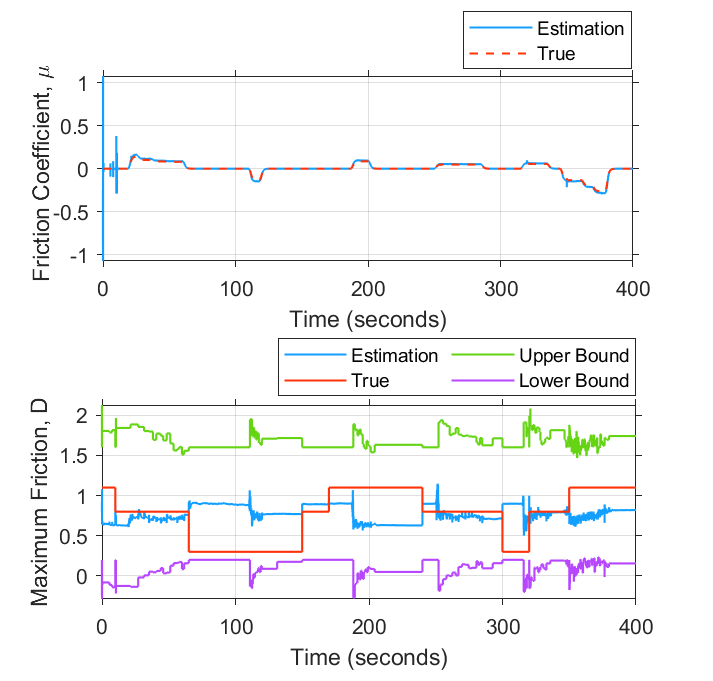}
    \caption{Identification of both current friction coefficient (top) and maximum available friction (bottom) over a varying road surface shown by the change in true maximum friction coefficient (red). 
    %Also shown is the Particle Filter particle range shown by the maximum (green) and minimum (purple) particle values.
    Note that the blue curve is the mean value of the estimation, while the green and purple curves show the upper bound and the lower bound of the estimation.}
    % \textcolor{red}{explain the meaning of the lines e.g. maximum, minimum and estimate explicitly. For example, estimate is calculated as the mean of the distribution of the friction estimation }}
    \label{fig:passive}
\end{figure}
One may argue that an estimator (a particle filter for example) may in fact be able to eventually identify a parameter whilst only requiring the passive stimulus from the driver of a road vehicle \emph{e.g.} throttle or brake input. 
However separating the belief update and control action selection does not always perform accurate identification as shown in Fig. \ref{fig:passive} where the road surface friction coefficient and parameter D (maximum road friction) of the Magic formula are estimated over a changing road surface. This uses the particle filter and simulation vehicle discussed in the remainder of this paper.
The particle filter represents its belief using a distribution of particles and associated weights that are updated upon receiving a new measurement. The estimation is therefore the weighted mean of the distribution.
We can see that the current friction coefficient is estimated well after a short identification period.
However parameter \(D\), vital in the emergency braking procedure, is not estimated well due to the fact that the vehicle hasn't adequately explored the state space.

% Notably driving the vehicle to the maximum tyre force substantially improves the identification of this parameter.
In this example the stress (or uncertainty) of the observer is shown using the range of the particle distribution (green and purple lines) that never improves toward the expected value shown in blue. Convergence of the particles is expected if the estimator is behaving as intended.
In this regard, given the uncertainty is very high, no control policy could be confidently derived from it. 
% Furthermore, the estimation frequently transitions between estimation periods that are stable and unstable.
Thus, the motivation for actively choosing a control policy that explores the agent state and parameter space is necessary, taking into account the belief, hence an active learning approach, such as DCEE is necessary.

\section{Torque Vectoring for Active Learning}\label{sec:4}
% This scheme guarantees that understeer will happen first during TV thus is safe if operation drives vehicle to become unsafe.
% The motivation for using an active learning method over a passive approach is demonstrated and highlights the motivation for TVAL over traditional approaches. Then we go onto to present the TVAL control scheme.

Identification of the peak friction coefficient is achievable when using the exploration-exploitation nature of DCEE \emph{e.g.} as part of the Anti-lock Brake System (ABS) in \cite{sullivan2023}. 
However, when planning if emergency braking is required, both the peak and current friction coefficients must be known \emph{a-priori}. 
Periodically applying ABS is not possible since extreme braking is not acceptable as part of normal, driver controlled, driving.
Thus we present a method using the DCEE framework \cite{Chen2021} to achieve this.
The control scheme can be divided into two parts, firstly an observer which updates the belief and performs the identification of the partially observable states and parameters. 
This is briefly discussed in this section although full description can be found in \cite{sullivan2023}. 
Secondly the whole vehicle torque vectoring for active learning  (TVAL) is introduced that updates the action in an exploration-exploitation fashion whilst following the driver’s reference.

\subsection{Torque Vectoring Control Strategy}
This work assumes that states and parameters of the Magic Formula tyre model (\ref{eqn:magic}) are unknown and should be estimated in an online manner. 
A Regularized Particle Filter observer is used to generate a belief in the augmented state given sequential sensor observations of the vehicle states.
This provides the necessary level of accuracy without offline training and is more beneficial than those using Deep Learning \cite{Li2020} for example.
The system dynamics is described in (\ref{eqn:sysmod}) using the vehicle state \(\mathbf{x}\) that are the vehicle body velocity \(v\) and the front and rear axle wheel speeds \(\omega_{f,r}\) \emph{i.e.} \(\mathbf{x}=\left[ v,\omega_{f},\omega_{r}\right]^T\). 
In addition,  \(\mathbf{u}\) represents the control input for each wheel, \emph{i.e.}, brake or throttle, while  \(\mathbf{\Theta}=\left[B,C,D,E \right]^T\) denotes the parameters of the tyre model. %where \(\mathbf{\Theta}=\left[B,C,D,E \right]^T\). 
No lateral motion of the vehicle is assumed and hence only braking or driving is considered.
\begin{equation} \label{eqn:sysmod}
\begin{aligned}
    \mathbf{x}_{k} & = f\left(\mathbf{x}_{k-1},{\mathbf{u}}_{k-1},\mathbf{\Theta}_{k-1} \right) \\  
    \mathbf{y}_{k} &= \mathbf{x}_{k} + \mathbf{V}_{k}
\end{aligned}
\end{equation}

\begin{remark}
    %Although some researchers consider lateral and longitudinal behaviour, 
    % AEB is predominantly a longitudinal feature. 
    % Compared with other driving maneuvers containing both lateral and longitudinal behaviour, less information could be used to learn the unknown parameters, making the identification task more challenging in practice. 
    % During normal driving lateral vehicular accelerations are much higher than forward acceleration hence the contribution from the passive learning effort is greater since larger tyre forces are generated.
    During normal driving, lateral vehicular accelerations are higher than forward accelerations hence the state space can be explored more using just a passive input.
    Therefore the longitudinal only motion that we consider here presents a more challenging scenario to perform identification.  
\end{remark}

The measurements of the vehicle states, \(\mathbf{y}\), have independent additive Gaussian noise, \(\mathbf{V}\). The augmented state matrix is defined as the states of the vehicle, \emph{i.e.}, body and wheel speeds and all four Magic formula parameters:
\begin{equation}\label{eqn:abssystdyn}
\mathbf{\chi}_k = 
    \begin{bmatrix}
        \mathbf{x}_{k}\\
        \mathbf{\Theta}_{k}
    \end{bmatrix}
    =
    \begin{bmatrix}
        f\left(\mathbf{x}_{k-1},\mathbf{u}_{k-1},\mathbf{\Theta}_{k-1} \right)\\
        \mathbf{\Theta}_{k-1}
    \end{bmatrix}
\end{equation}
% The Regularized Particle Filter is shown in \ref{alg:rpf}.
% \begin{algorithm} %{0.6\textwidth}
% \caption{Regularized Particle Filter with MCMC Step}\label{alg:rpf}
%  \hspace*{\algorithmicindent} \textbf{Input:}  \(\mathbf{\chi}_{k-1}^i,w_{k-1}^i,\mathbf{y}_k \)\\
%  \hspace*{\algorithmicindent} \textbf{Output:} \(\mathbf{\chi}_{k}^i,w_{k}^i \)
% \begin{algorithmic}[1]
%     \STATE Draw Particles using  (\ref{eqn:sysmod})
%     \STATE Calculate likelihood \(w^i_k=p\left(\mathbf{y}_k \mid \mathbf{\chi}_{k}^i \right)\)
%     \STATE Find the effective ratio \(N_{eff} = 1/\sum_{i=1}^{n_i}\left(w_k^i\right)^2\)
%     \IF{\(N_{eff} \leq K_{o,k} n_i\)}
%         \STATE Systematic Particle Resampling 
%         \STATE Regularization Step
%         \STATE Markov Chain Monte Carlo (MCMC) Step
%     \ENDIF
%     \end{algorithmic}
% \end{algorithm}
% Following the observer \(n_i\) particles and weightings can be used to find the expected value of the augmented state vector at each sampling interval using (\ref{eqn:exp}).
% \begin{equation}\label{eqn:exp}
%     \mathbb{E} \left[\mathbf{\chi}_k\right] = \sum_{i=1}^{n_i} W_k^i \mathbf{\chi}_k^i
% \end{equation}

A simplified discrete time 3DOF, longitudinal model is used here for estimation and prediction where we account for front and rear wheels along with the normal load that we assume to be known \emph{a-priori}. Since there is no lateral load transfer, the right and left wheels on each axle produce the same tyre force.
Using the Forward Euler Method, the predicted ego body velocity is found as:
\begin{equation}
    {v}_{k} = \frac{2 \Delta t}{m} \left[ {F}_{k-1,f} + {F}_{k-1,l}\right] + {v}_{k-1}
\end{equation}
Similarly each wheel speed can be calculated by 
\begin{equation}
    {\omega}_{k,\{f,r\}} = \frac{\Delta t}{I_{w}}  \left[ u_{k-1,\{f,l\}} - {F}_{k-1,\{f,r\}} R \right] + {\omega}_{k-1,\{f,l\}}
\end{equation}
where \(\Delta t\) is the sampling period and we assume that the wheel inertia is the same between each wheel.
The parameters are assumed to be piece-wise constants, \emph{i.e.}, \(\mathbf{\Theta}_k=\mathbf{\Theta}_{k-1}\). Then using (\ref{eqn:tforce})-(\ref{eqn:slip}) the predicted tyre force can be found using the vertical load on the tyre.
% \begin{remark}
% The assumption that the parameters remain constant is valid since we initialize the particle population in a way to cover the vast majority of tyre models. Furthermore, the weather changes at a very slow frequency compared to the rate of the observer and the changes in weather and tyre model are modelled as constant as has been done previously.
% \end{remark}

Using the Particle filter formulation from \cite{sullivan2023}, the augmented state matrix is estimated at every discrete time step \(k\) using \(i\) number of particles:
\begin{equation}\label{eqn:exp}
    \bar{\chi}_k = \mathbb{E} \left[\mathbf{\chi}_k\right] = \sum_{i=1}^{n_i} W_k^i \mathbf{\chi}_k^i
\end{equation}
In this work we use \(10,000\) particles although this may be tuned to trade-off between computational speed and accuracy.

The DCEE framework \cite{Chen2021} acknowledges that a future action will change the future observation. Hence using the aforementioned system model one may predict the observation of the vehicle's tyre forces. For efficiency, we hypothesise over a finite set of discrete actions \(\mathcal{T}\) and allows for us to incorporate the rate limitations of the actuator dynamics into the control policy. In this case we use the following action set; \(\mathcal{T}\in \left[0,\pm1,\pm10,\pm100\right]\).
The use of electrified powertrains allows the application of torque to be achieved much faster and more accurately \cite{Heerwan2016} than traditional vehicles. Furthermore the use of individual motors for each wheel is more effective than active differentials which is traditionally the case \cite{DeNovellis2014}.
This makes the use of torque vectoring and the selection of our actions realistic and realisable for improving the stability of vehicles. 

% Furthermore  Four-wheel independently actuated electric vehicles (FWIA EVs) are consider to be the ideal case since all wheels are available to control the vehicle as opposed to front or rear wheel drive only.
TVAL aims to drive the vehicle to explore the epistemic uncertainty of the maximum tyre at one axle, then using the wheels at the other axle to provide a reactive force and achieve the acceleration demand from the driver. The objective for the vehicle can be expressed as:
\begin{equation}\label{eqn:vehgoal}
    ma_{ref}=\sum^{4}_{j=1} F_{xj}
\end{equation}
where \(a_{ref}\) is the driver acceleration demand from the brake and throttle inputs.
Thus, this work aims to provide an autonomous means to explore the road surface without interfering with the driver.
However our goal orientated approach is to explore the maximum limits of adhesion thus we do not use (\ref{eqn:vehgoal}) as the subject of our control scheme.
Instead, we aim to find a policy that drives the rear wheel towards the maximum tyre force that is unknown but can be predicted. 
Additionally, due to the dual nature of this controller, the uncertainty of the tyre being at the maximum tyre force is also evaluated in the same optimization framework.
We therefore reduce the cognitive dissonance of the system in an online manner by giving action to the perception and reducing uncertainty.

The optimization routine is outlined in
(\ref{eqn:aebopt}) where the objective is to track the predicted future tyre force at the rear wheel, \(F_{k+1 \mid k,r}^*\). This is estimated using the posterior estimate of the maximum friction coefficient (that is parameter \(D\) from the Magic Formula tyre model).
The predicted friction coefficient \({\mu}_{k+1\mid k,r}\) can then be used similarly with the vertical tyre load to find the predicted tyre force \(F_{k+1 \mid k,r}\).
Since we use predicted observations to inform the likelihood of predicted states, predicted tyre force measurements are generated using the aforementioned system model and the future expected state found using (\ref{eqn:exp}).

\begin{subequations}\label{eqn:aebopt}
\begin{align}
    \min _{\mathbf{U}_{tv,k}} J\left(\mathbf{U}_{tv,k}\right)&=  \min_{\mathbf{U}_{tv,k}} \mathbb{E}_{\mathbf{\Theta}_k}\left[\mathbb{E}_{y_{k+1|k}}\left[\big|F_{k+1 \mid k,r} -F_{k+1 \mid k,r}^*\big| \mid \bar{\mathbf{\chi}}_k,\mathbf{y}_k\right]\right]\\
% \min J\left(\mathbf{u}_k \right) &= \min_{\mathbf{u}_{k}} \big| {F}_{k+1 \mid k,f} - {F}_{k+1 \mid k,f}^*\big| + {S}_{k+1 \mid k,f}  \\
\text{ Subject to: } \notag \\
\text{ \underline{DCEE Wheel} } \notag \\
{F}_{k+1 \mid k,r} &= {\mu}_{k+1 \mid k,r} F_{k+1 \mid k,rz} \\
{F}_{k+1 \mid k,r}^* &= {D}_{k+1 \mid k}  F_{k+1 \mid k,rz} \\
F_{k+1 \mid k,rz} &= F_{ k,rz} \\
{\Theta}_{k+1 \mid k} &= {\Theta}_{k} \\
% &{\mu}_{k+1\mid k,d} = Magic \left( {\kappa}_{k+1 \mid k,d},{\Theta}_{k+1 \mid k}\right)\\
% &{\kappa}_{k+1 \mid k,d} = \frac{{\omega}_{k+1 \mid k,d} R - {v}_{k+1 \mid k}}{{v}_{k+1 \mid k}} \\
% &{\omega}_{k+1 \mid k,d} = \left[ u_{k,d} - {F}_{k,d} R \right] \frac{\Delta t}{I_{w}} + {\omega}_{k,d} \\
% &{v}_{k+1 \mid k} = \frac{dt}{m} \left[ {F}_{k,a} + {F}_{k,d}\right] + {v}_{k} \\
{u}_{k,d} &= {u}_{k-1,d} + {\tau}_{k} \label{eq:aebopt1}\\
{\tau}_{k} &\in \mathcal{T} \label{eq:aebopt2} \\
\text{ \underline{Reactive Wheel} }\notag\\
% &\mu_{f,ref}F_{z,f} = \left[ m a_{u} - \mu_r F_{z,r} \right]\\
% &{\mu}_{k+1\mid k,a} = Magic \left( {\kappa}_{k+1\mid k,a},{\Theta}_{k+1\mid k} \right)\\
% &{\kappa}_{k+1\mid k,a} = \frac{{\omega}_{k+1\mid k,a} R - {v}_{k+1\mid k}}{{v}_{k+1\mid k}} \\
% &{\omega}_{k+1\mid k,a} = \left[ u_{k,a} - {F}_{k,a} R \right] \frac{\Delta t}{I_{w}} + {\omega}_{k,r} \\
% &{F}_{k,a} = {\mu}_{k,a} F_{k,az}\\
{F}_{k+1\mid k,f} &= {\mu}_{k+1\mid k,f} F_{k+1\mid k,fz} \\
\mathbf{U}_{tv,k} &= \left[u_{k,a},u_{k,d} \right] \\
\text{ \underline{Driver Intention} } \notag \\
F_{k,ref} &= a_{k,ref} F_{k,fz} \\
%\text{ Subject to: } \notag \\
F_{k+1\mid k,f}\left(u_{f} \right) &= F_{k,ref} - {F}_{k+1\mid k,r}\left(u_{k,r} \right)
\end{align}
\end{subequations}
% This scheme alters the active learning driven wheel and (D) and reactive wheel (A) according to the following condition:
% \begin{equation}\label{eqn:wdistconst}
%     F_{zf} < F_{zr}
% \end{equation}
% This is required since dynamic loading causes the load over each tyre to change leading to changing location of where the peak tyre forces are achieved.
% \textcolor{red}{add more explanation of the physical meaning of the optimisation problem such as the cost function; also noted physical meaning of $F^*$ is not defined}
Note that we make the same assumptions as we did for the observer that the parameters are piecewise constants. 
Furthermore we also assume that the vertical tyre load doesn't change over the one step prediction.

\subsection{Sensor Driven Resampling}
Perpetuated operation of the observer, as is the case in real world systems, leads to belief perseverance in the estimated state which is detrimental to sudden changes in the operating point. Similar concepts include perceptual narrowing in neuroscience disciplines. 
In other words, the believed uncertainty in the identification of the state estimation becomes exceptionally low when the observer has generated a very high belief, and  the estimation space converges to a very small region.
Using resampling based upon uncertainties therefore becomes degenerate if the road surface does not change for an extended period and the passive stimulus from the driver is small.
This behaviour is clear by looking at Fig. \ref{fig:muslip} since at low slip ratios it is challenging to distinguish between which road surface the vehicle is travelling on. 
This phenomenon led to the Retrogressive resampling procedure, introduced in previous work \cite{sullivan2023}.
% Different to previous Retrogressive resampling applications, this work requires the observer to provide estimations over significantly longer periods of time.
% Compared to (??) that demonstrated short  estimation period (\emph{i.e.}, less than $4 [s]$) over a single road surface change, this work requires accurate augmented state estimation over 400s, (100 times longer) on multiple road surfaces.
% Compared to \cite{sullivan2023} that demonstrated short estimation period (\emph{i.e.}, less than $4s$) over a single road surface change, this work however acknowledges that normal journey times are significantly longer and over different road surfaces.

The resampling procedure is required to avoid immediate degeneracy upon changing to different road surfaces. 
Motivated by the idea of using commonly available vehicular sensors, we use the rain and light sensor found on most passenger vehicles.
This sensor, typically located behind the rear-view mirror, is used as a driver assistance system to activate or adjust the frequency of the windscreen wiper, leaving the driver to focus on the road. 
Although these group of sensors may be viewed as rudimentary, other sensors such as humidity or temperature sensors may also be used to recognize a change in road surface however sophisticated vision based systems have also found practical application \emph{e.g.} \cite{Tian2021}.
We therefore may correlate the detection from the rain/light sensor to a changing road surface and the accumulation of rain or snow on the road surface.
\begin{remark} % quickly changing environments dont impact the controller, only power consumption.
Since the estimation time of the augmented state is very quick, the sensitivity on the controller to frequently changing environments is not significant. 
However this of course does come at the expense of any existing belief in the tyre and environment which must be re-identified at the cost of vehicle battery power.
\end{remark}

We can therefore modify the Retrogressive resampling algorithm to reset the parameter estimation to their initial, diverse state. This is shown in Algorithm \ref{alg:retro}.
\begin{algorithm} %{0.6\textwidth}
\caption{Retrogressive Resampling}\label{alg:retro}
 \hspace*{\algorithmicindent} \textbf{Input:}  \({\mathbf{x}}^i_k,\mathbf{\Theta}_0^i,\mathbf{\Theta}_k^i,w^i_0,w_{k}^i,\rho_k,\rho_{k-1}\)\\
 \hspace*{\algorithmicindent} \textbf{Output:} \( {\mathbf{x}}^i_k,\mathbf{\Theta}_k^i,w_k^i\)
\begin{algorithmic}[1]
    \IF{\(\rho_k \neq \rho_{k-1}\)}
        \STATE \( \mathbf{\Theta}_k^i \gets \mathbf{\Theta}_0^i\)
        \STATE \({\mathbf{x}}_{k}^i \gets \mathbf{x}^i_{k}\)
        \STATE \({w}_{k}^i \gets w^i_0\)
    \ELSE
        \STATE \( \mathbf{\Theta}_k^i \gets \mathbf{\Theta}_k^i\)
        \STATE \({\mathbf{x}}_{k}^i \gets \mathbf{x}^i_{k}\)
        \STATE \({w}_{k}^i \gets w^i_k\)
    \ENDIF
\end{algorithmic}
\end{algorithm}

\section{Torque Vectoring Demand Regulation}\label{sec:5}
Continuous active learning is inefficient, unnecessary, and potentially unsafe. Here the availability, energy consumption and safety of performing TVAL is considered. Additionally, a smooth integration scheme between active and passive (driver only) is required.

\subsection{Energy Consumption}
\begin{figure}
    \centering
    \includegraphics[scale=0.72]{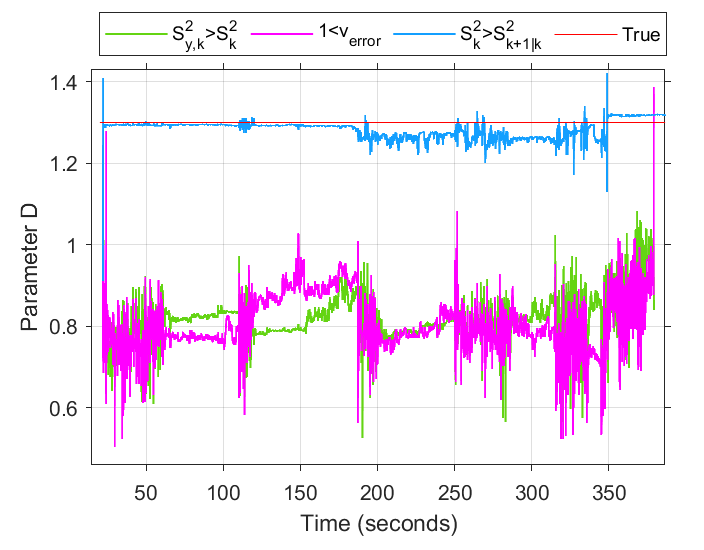}
    \caption{Comparison of different TVAL activation strategies. Results using sensor \(S_{y,k}^2\) and estimation \(S_{k}^2\) uncertainty are shown in green, error between sensor measurement and estimation is compared against a threshold of \(1\) in pink and the result using estimation and prediction uncertainty is shown in blue.}
    \label{fig:TVALacti}
\end{figure}

Torque vectoring requires significant energy since it requires energy to perform additional tasks above those required to drive or brake the vehicle. 
Hence management of resource consumption must be addressed particularly since consumers place high weighting on battery range. 
One method to quantify this would be to compute the work done by each wheel and find the energy consumption above that required by the driver then defining a minimization policy. 
However, this is not a goal orientated approach to resource management, rather solely a resource management approach which is unaware of the identification goal of the controller. 
If the system can generate a belief of high confidence in the friction coefficient and associated parameters, then it does not need to continuously operate since we may not be able to improve the estimation further.
A comparison of different activation thresholds based on measurement and estimation error, \(v_{error}\) (pink line), sensor accuracy \(S^2_{y,k}\) and estimation uncertainty \(S^2_{k}\) (green line), and estimation and prediction uncertainty \(S^2_{k+1\mid k}\) (blue line) is presented in Fig. \ref{fig:TVALacti} while estimating parameter D over \(400s\) using the Extra-Urban Drive Cycle (EUDC).
It is clear to observe that the most effective way to activate TVAL is to compare the estimation uncertainty with the predicted one calculated using 
\begin{equation}
    S_{k+1 \mid k}^2 = Var \left( \bar{F}_{k+1\mid k}^* \right)
\end{equation}
%This can also be regarded as the uncertainty of the  estimated maximum tyre force, substituting the predicted tyre forces for the estimations.
%From a system perspective this approach is most logical since we can exploit the belief of the predicted measurements generated within the DCEE framework.
When the uncertainty in the maximum force prediction becomes less than that in the estimated maximum force, 
the system is required to improve the estimation through the exploration and exploitation effort. 
This can be achieved using the covariances \(S^2\) and the following relationship:

\begin{equation} \label{eqn:power}
    \tau_p = S^2_{k} > S^2_{k+1 \mid k}
\end{equation}
which when satisfied, requires the DCEE to be activated in learning the new system states and thus improving the estimate of augmented state matrix.

\begin{figure}
    \centering
    \includegraphics[scale=0.5]{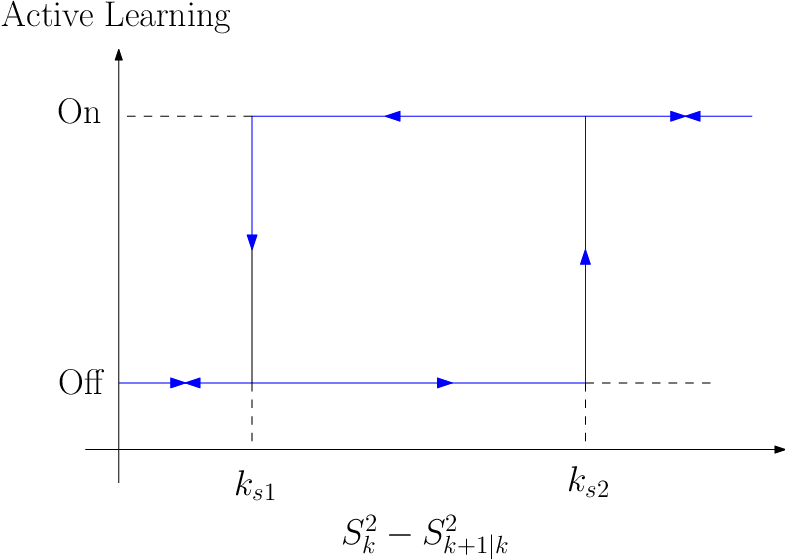}
    \caption{Energy saving Hysteresis switch design. The error between uncertainties on predicted and estimated maximum tyre forces is used to switch between On and Off active learning states. Two gains, \(k_{s1}\) and \(k_{s2}\) are used to govern the behaviour. }
    \label{fig:hystswitch}
\end{figure}

However, this can result in unnecessary chattering around the threshold that does not improve the estimation. 
Hence although a reduction in the uncertainty caused by a selected action may temporarily reduce the uncertainty, it may not have reached a steady state. 
% \begin{figure}
%     \centering
%     \includegraphics[scale=0.87]{Images/EUDC_switchnohyst.eps}
%     \caption{Switching behaviour using threshold method where \(1\) and \(0\) denotes the TVAL being activated or deactivated respectively. ( REMOVE FIGURE???)}
%     \label{fig:switch}
% \end{figure}
Therefore we employ a Hysteresis switch to manage the activation of the TVAL scheme, as shown in Fig.~\ref{fig:hystswitch}. Here the reset point \(k_{s1}\) determines when TVAL is no longer required and driver only operation is requested. 
Then a switching point \(k_{s2}\) indicates the threshold where the system is suitably uncertain or stressed and hence the TVAL is activated.

\subsection{Vehicle Stability}
Vehicle understeer or oversteer occurs when less steering (oversteer) or more steering (understeer) is required by a vehicle to remain on a given curved path, as its forward speed increases.
In the case of understeer, the vehicle is said to turn less than the reference input and conversely an oversteering vehicle turns more. 
In our longitudinal only problem, this remains an important consideration since the high forces experienced by the tyre during torque vectoring reduces the responsiveness to steering inputs.
This instability can be estimated using the understeer gradient and is dependent on the cornering stiffness of the front (and rear) tyres $C_{\alpha f}$ (and $C_{\alpha r}$) as
\begin{equation} \label{eqn:undsteer}
    K_{us} = \left( \frac{F_{zf0}}{C_{\alpha f}} - \frac{F_{zr0}}{C_{\alpha r}} \right)
\end{equation}%shown in (\ref{eqn:undsteer}).
where \(F_{zf0} \) (and $F_{zr0}$) represents the static load at the front (and rear) axle \cite{Pacejka2002}. 
$K_{us}$ represents the understeer gradient. If \(k_{us}>0\) then the vehicle is said to be understeering and if \(k_{us}<0\), then the vehicle is oversteering.
Although both phenomena are undesirable, oversteer is unstable and considered to have greater risks since advanced driver behaviours are required to regain control. Conversely, escaping understeer is achieved by reducing the vehicle velocity.
The critical speed \(v_{crit}\), used to determine at what speed oversteer occurs, is calculated by 
\begin{equation} \label{eqn:vcritical}
    v_{crit} = \sqrt{\frac{gl}{\mid K_{us} \mid}},
\end{equation}
where $g$ is the gravity constant.
%  and $l$ denotes {\color{red}XXX} % l is already defined
%It is used to find at which point the vehicle starts to spin.
This allows us to compare current or predicted ego vehicle velocities to the critical velocity and determine instability in an online fashion.
According to \cite{Pacejka2002} we can find the individual cornering stiffness as a function of the friction coefficient \(\mu_k\), vertical \(F_{k,z}\) and longitudinal \(F_{k,x}\) loads for each wheel:
% \begin{equation}
\begin{equation}
C_{k,\alpha}\left(\mu_k,F_{k,z},F_{k,x}\right)=
\varphi_{k,x \alpha}\left[ C_{k,\alpha} \left(F_{k,z}\right)
-\frac{1}{2} \mu_k F_{k,z}\right]
+\frac{1}{2}\left(\mu_k F_{k,z}-F_{k,x} \right)
\end{equation}
% \end{equation}
where
\begin{equation}
\varphi_{k,x \alpha}=\left[1-\left(\frac{F_{k,x}}{\mu_k F_{k,z}}\right)^n\right]^{1 / n}
\end{equation}
using \(n\) between \(2-8\). Additionally, the cornering coefficient as a function of the vertical load (only) is found using the coefficients \(c_1\) and \(c_2\) (used to define the cornering stiffness) and nominal load \(F_{z0}\):
\begin{equation}
C_{k,F \alpha} = c_1 c_2 F_{z o} \sin \left\{2 \arctan \left(\frac{F_{k,z}}{c_2 F_{z o}}\right)\right\} 
\end{equation}
Therefore to enforce the stability, the following constraint should be applied: %we can define an additional constraint to limit the behaviour of the exploration effort which should not operate if the ego vehicles velocity is above the critical speed, \emph{i.e.},:
\begin{equation} \label{eqn:vcrit}
    \tau_{k,s} = v_{k+1 \mid k} < v_{k+1 \mid k,crit}
\end{equation}
% The difference between our work and others is that others often consider the stability of tracking the current reference \emph{e.g.} does the predicted lane change reference result in oversteer. 
% However, we consider that the future state should not be one that has the potential to become unstable.
%Whilst one approach may be to consider estimating if the vehicle is currently oversteering, 
Note that we consider predicting if the vehicle has the \textit{potential} to oversteer.
Since torque vectoring places additional demand on the vehicle, it is important to recognize that the drivers input can be unpredictable. 
Hence the above constraint is necessary for (partially) autonomous systems such as automated driving.

\subsection{Tyre Force Availability}
% This section introduces a constraint to limit when TVAL is used. Since the request is not always possible.
%

If the maximum friction coefficient is known, or in this case estimated, the vehicle's maximum permissible acceleration across any road surface can be calculated. 
With this information available to our controller, we can assess whether the driver's acceleration request is achievable while simultaneously identifying the augmented state vector.
% This idea is shown in Fig. \ref{fig:avtyre} where the acceleration and deceleration limits are shown when performing active learning (coloured circles) and when the vehicle is operating without (dashed) \emph{i.e.} driver only. 
This limitation is found as a result of the differential loading between the front and rear axles of the vehicle. 
If the vehicle were to be excessively front loaded \emph{i.e.} the majority of the vehicle's weight were over the front axle, then a much greater acceleration would be possible. 
Since there would be a smaller maximum tyre force at the rear required to be explored (for the purposes of actively learning the road surface) and therefore much more tyre force at the front to equal this and also achieve the acceleration request from the driver.
% In Fig. \ref{fig:avtyre} the solid black line shows the tyre force required for just achieving the desired acceleration.
% We can see that TVAL comes at the expense of not being able to achieve the driver’s request. 
% For example, if the driver wanted to accelerate at \(1m/s^2\), TVAL would only be achievable on a dry road surface. Yet without TVAL, the vehicle could achieve this on any road surface discussed here as shown by the coloured dashed lines. 
% While decelerating, only the friction limits of the road surface constrain the controller.
% One may argue that TVAL is too limited to work within the operation of normal driving behaviours. This however is not the case when considering drive cycle descriptions of \emph{human} behaviour. 
% In the case of the maximum accelerations found in the EUDC (demonstrated in Section~\ref{sec:6}) and the WLTP (Worldwide Harmonised Light Vehicles Test Protocol), the vast majority of driving allows for the use of TVAL, depending on the surface. 
% The peak acceleration found in each of these cycles is shown on Fig \ref{fig:avtyre} as dash-dotted and dashed lines repectively. 
% Furthermore, a singularity exists where the majority loaded axle switches from the front (positive tyre force) to the rear. In this simple model which uses a simple half car model with data from Section~\ref{sec:6}, an acceleration of \(4.43m/s^2\) causes this vanishing point although exists outside normal driving behaviour.
Hence a constraint (\ref{eqn:ava}) is chosen to govern the time varying activation and deactivation of TVAL considering the maximum availability of tyre force using the estimated friction coefficients \(\bar{F}^*_{fx}\). %This constraint limits the activation of TVAL when the tyre force required to do active learning and achieve the driver input does not exist.
\begin{equation} \label{eqn:ava}
    \tau_{a,k} = \frac{m}{2} a_{ref,k} \geq \bar{F}^*_{fx,k} + \bar{F}^*_{rx,k}
\end{equation}

\begin{remark}
    This further highlights the need for active learning, since only at the extreme acceleration of \(\pm10.8m/s^2\) would the driver exploit the full friction of the dry road.
    Our approach can operate at much lower, typical, vehicle accelerations. In fact it can perform at zero acceleration, i.e. cruising.
\end{remark}

\subsection{Input Management}
% \textit{comfort} hysteresis is introduced to successfully integrate all constraints into the TVAL framework which benefits the controller in the following ways. First, since their may be a residual affect of TVAL on the vehicle occupants, hysteresis is required to mitigate this effect caused from rapidly changing road surfaces. 
% This rejects small changes in the road surface from causing TVAL to become active.
% It additionally prevents chattering from the combined effort of each constraint, forcing a pause between torque vectoring identification effort.
% Here, this is implemented using the period between activation cycles. For example, one may introduce a 5 second delay between successive identification efforts.

\begin{figure}
    \centering
    \includegraphics[scale=0.6]{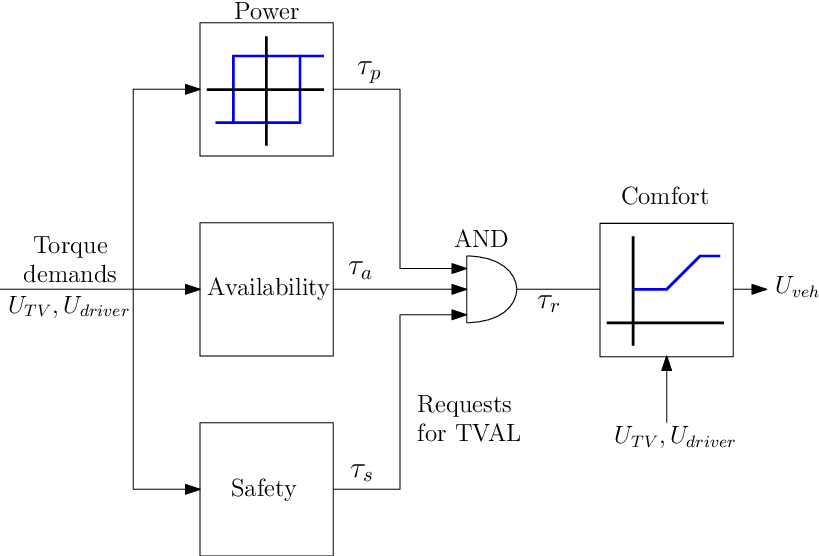}
    \caption{Activation modulation scheme showing the mapping of 3 constraints and controls  to facilitate the transition between driver and active learning.}
    \label{fig:heihyst}
\end{figure}
The request for active learning is simply if all Eqs. (\ref{eqn:power}), (\ref{eqn:vcrit}) and (\ref{eqn:ava}) are satisfied. %\emph{i.e:}
%\begin{equation}
    %\tau_{k,r} = \tau_{k,p} \land \tau_{k,a} \land \tau_{k,s}
%\end{equation}
Switching on or off between active learning and driver controls, after the tyre model has been identified and TVAL constraints satisfied is not a suitable strategy since the sudden deactivation could lead to front wheel torque inputs to rapidly decrease or increase. 
This leads to an increase in acceleration and thus greater discomfort on the driver.
\begin{remark}
    This highlights the challenge associated with higher level autonomous driving such as SAE level 3 autonomous driving \cite{J3016_202104}. 
    Changing between driver and autonomous control cannot be as straightforward as on or off and must be managed carefully to ensure smooth, controllable handover \cite{Hu2024}.
\end{remark}

To mitigate this and control the learning activation rate, a linear mapping between the two controllers is now presented, that is, between the driver and our torque vectoring scheme. This is shown in  (\ref{eqn:linmapweight}) using the wheel inputs as calculated by the torque vectoring controller, \(\mathbf{U}_{tv,k}\), and the driver input \(\mathbf{U}_{driver,k}\): 
\begin{equation} \label{eqn:linmapweight}
    \mathbf{U}_{veh,k} = w_{1,k} \mathbf{U}_{tv,k} + \left(1 - w_{1,k}\right) \mathbf{U}_{driver,k}
\end{equation}
Using weight \(w_{1,k}\) to moderate the transition between the two systems.
This rejects small changes in the road surface from causing TVAL to become active.
It additionally prevents chattering from the combined effort of each constraint, forcing a pause between torque vectoring identification effort.
Simply, if the request for active learning (\(\tau_{r,k}==1\))  is satisfied most of the time, then torque vectoring should become activated. Conversely, if there is no demand for active learning \emph{i.e.} \(\tau_{r,k}==0\) is satisfied most of the time, then only the drivers input should be passed to the vehicle. 
This request counter relationship, \(f_k\) is captured using equation (\ref{eqn:linmapfn}):
\begin{equation} \label{eqn:linmapfn}
    f_k = f_{k-1} + \left(\tau_{r,k}==1\right) - \left(\tau_{r,k}==0\right).
\end{equation}
To ensure that the weighting is bound between \(0\) and \(1\), the request counter function is bounded as well:
\begin{equation} \label{eqn:hystf}
    f_{k} = \begin{dcases}
        0, & f_{k}=0 \\
        f_{k}, & 0 < f_{k} < \Delta P \\
        \Delta P, & f_{k} \geq \Delta P
    \end{dcases}
\end{equation}
Finally, the weighting function can be found using the request gradient, and \(\Delta P\) is tuneable depending on the desired transition period. The tuning methodology follows that one should select \(\Delta P\) to be the transition time between either inputs which may be in the order of a few seconds.
Finally, the time dependant weighting can be found as:
\begin{equation}
    w_{1,k} =
    \frac{f_k}{\Delta P}
\end{equation}
The modulation of the active learning is shown in Fig. \ref{fig:heihyst} that is a combination of all three constraints which drives the application of the driver and torque vectoring inputs (\ref{eqn:linmapweight}) to the vehicle.
% Using the active wheel indicators where \(A_{ct}=0\) if the front wheel is reacting to the DCEE control input and  \(A_{ct}=1\) if it is at the rear. In other words if the relationship given in (\ref{eqn:wdistconst}) is met then the 
% \begin{equation}
%     A_{f,k} = \left(A_{ct}==0\right) - \left(A_{ct}==1\right)
% \end{equation}
% \begin{equation}
%     A_{r,k} = \left(A_{ct}==1\right) - \left(A_{ct}==0\right)
% \end{equation}

% \begin{equation}
%     \Bar{D}_{k+1 \mid k}F_{fz,k+1 \mid k} \geq F_{fx,k+1 \mid k}
% \end{equation}
% \begin{equation}
%     \Bar{D}_{k+1 \mid k}F_{rz,k+1 \mid k} \geq F_{rx,k+1 \mid k}
% \end{equation}
% Conversely we can design the weight distribution of the car given what we expect the driver to do. If the driver doesn't exceed the \(\pm 1.389 m/s^2\) maximum acceleration, then using relationship for static load distribution one may find the location of the centre of gravity. Firstly from the rear axle \(b\),
% \begin{equation}
%     \frac{mgb}{2l}= F_{fz} = \frac{1}{D} \left(\frac{m}{2}a_{ref} - F_{rz} D \right) 
% \end{equation}
% and then from the front,
% \begin{equation}
%     a = l-b
% \end{equation}
% This ignores the the transient loading of the suspension.
% As an example using the EUDC maximum acceleration of \(\pm 1.389 m/s^2\) one may find that TVAL can successfully achieve the desired tyre forces if the weight distribution is \(59.8/40.2\) where \(a=1.22m\) and \(b=1.81m\). This is found using the length of a Jaguar XE vehicle.

% With the absence of gyroscopic sensors, we negate the impact the of the suspension here.
\section{Drive Cycle Performance} \label{sec:6}
\begin{figure}
    \centering
    \includegraphics[scale=0.7]{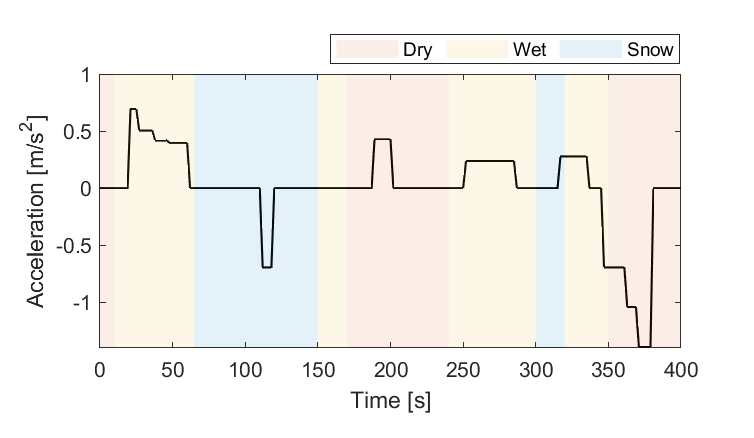}
    \caption{The EUDC acceleration profile across changing road conditions for dry (red), wet (yellow) and snow (blue) road surfaces.}
    \label{fig:setup}
\end{figure}

%
% SELCTIVE DISONANCE - chosing to oversteer
%
In lieu of comparative studies, we demonstrate our controller using a drive cycle and show that the energy and power consumption is significantly reduced whilst maintaining an accurate estimate of both current friction and maximum friction.
A changing road surface (see Fig. \ref{fig:setup}) is also tested to further show the ability to maintain a stable estimation and adapt to changes in the environment.
In this scenario we assume that a GPS and wheel speed senors are used with zero bias Gaussian noise \emph{i.e.} \( \mathbf{V}_k \sim \mathcal{N}\left(\left[0,0,0\right]^T, diag\left(0.2,0.5,0.5\right)\right)\). 
A driver model is used to simulate the driver and to track the reference acceleration profile. 
This is a PI controller that responds to the tracking error where a drive cycle provides the drivers acceleration reference over the entire simulation.
This model is shown in Fig. \ref{fig:SAD}.
The EUDC was chosen over the most recent WLTP since it has simpler behaviours thus as to not excessively passively perturb the system. 
This in turn requires our controller to actively explore the system more and hence shows a more challenging scenario.
The associated proportional and integral gains are \(k_P = 0.01\) and \(k_I = 15\) respectively. The Particle Filter is initialized with the properties in Table \ref{tab:init}.
% \begin{table}[h]
%     \caption{PI 'Driver' control properties.}
%     \centering
%     \begin{tabular}{|c|c|}
%     \hline
%         \multicolumn{2}{|c|}{PI Controller} \\
%     \hline
%         P Gain & 0.01 \\
%     \hline
%         I Gain & 15 \\
%     \hline
%     \end{tabular}
%     \label{tab:PI}
% \end{table}
\begin{table}[h]
    \caption{Particle Filter range of initial values.}
    \centering
    \begin{tabular}{|c|c|c|c|}
    % \hline
    %     \multicolumn{4}{|c|}{Observer Initial Conditions} \\
    \hline
        \(v_0\) & \(w_{f0}\) & \(w_{r0}\) & \(B_0\) \\
    \hline
        \(\left[ 0.5, 4.5\right]\) & \(\left[ 2.0, 9.0\right]\) & \(\left[ 2.0, 9.0\right]\) & \(\left[ 4.0, 21\right]\) \\
    \hline
    \hline
        \(C_0\) & \(D_0\) & \(E_0\) & \\
    \hline
         \(\left[ 1.3, 1.7\right]\) & \(\left[ 0.2, 1.6\right]\) & \(\left[ 2.0, -12\right]\) & \\
    \hline
    \end{tabular}
    \label{tab:init}
\end{table}

\subsection{Controller Stability}
\begin{figure}
    \centering
    \includegraphics[scale=0.75]{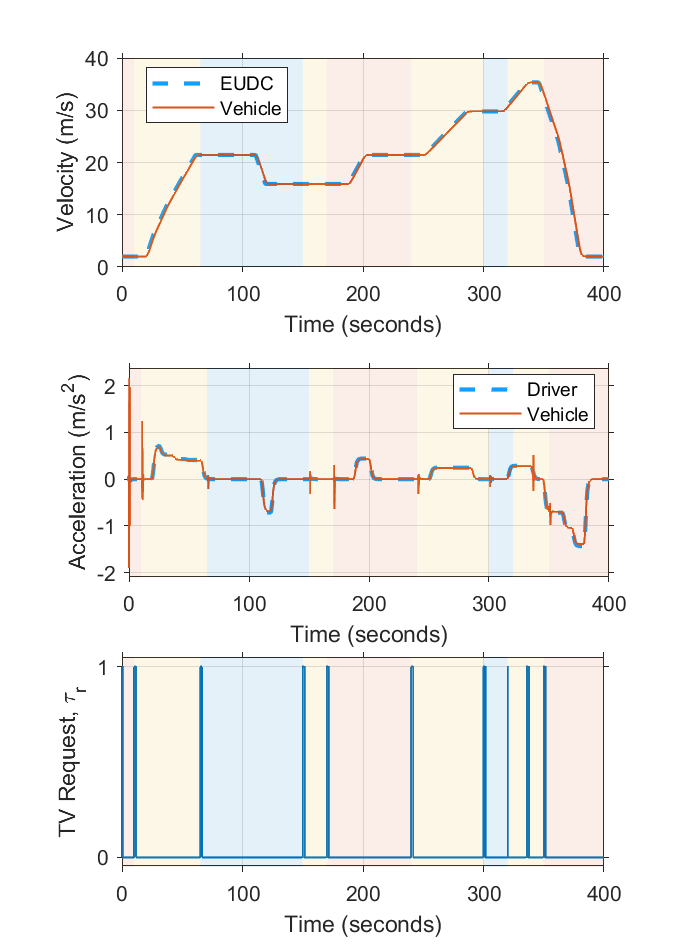}
    \caption{Vehicle velocity (top) compared with the EUDC (desired) reference, vehicle acceleration (middle) against the drivers acceleration request and TVAL request, \(\tau_r\) (bottom).}
    \label{fig:kineeudc}
\end{figure}

\begin{figure}
    \centering
    \includegraphics[scale=0.77]{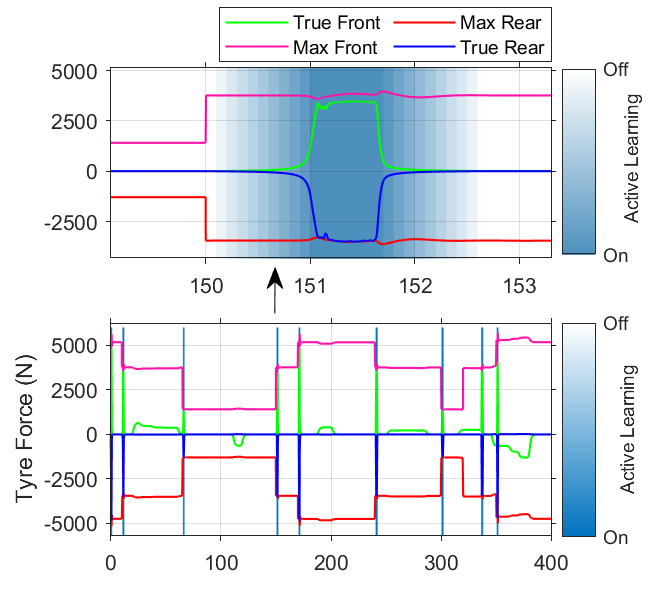}
    \caption{Front and rear vehicle tyre forces with the activation weight \(w_{1,k}\) overlaid as a colormap where dark blue represents full active learning input to the vehicle and white shows driver only. The maximum tyre forces for front and rear are also shown.}
    \label{fig:tyreforce}
\end{figure}

%y acceleration tracking
%y velocity tracking
%y prolonged estimation and velocity inaccuracy
% activation: period, high frequency oscillations
% yTyre forces: rising uniformly (fr), stable at max, rear tracks max well. Small oscillation at handover between 'driver' and TVAl
%y only at 300s and 330s does the tyre force not reach max (wet road, oversteer warning)
% torque at 150s shows oscillation (r), exploration, doesnt translate to large tyre force
Whilst the standard EUDC model has prolonged initial and final periods of stationary behaviour, we offset the velocity and initialize the vehicle starting at \(2m/s\). 
This is beneficial for demonstration since TVAL is not required when the vehicle is stationary.
It is assumed that the vehicle is running on an unknown road surface. Thus, it must actively probe the environment through TVAL to reduce the uncertainty in its belief of friction. A wide initial distribution of particles in our particle filter setting represents large uncertainty in beliefs of the augmented state. 
In Fig. \ref{fig:kineeudc}, TVAL is shown to activate immediately and follow a constant velocity on the initially dry road surface. 
% vel, accel, trigger
Overall the acceleration of the ego vehicle tracks that of the driver quite well, with deviations observed only during short transient periods when %with only transient increases either upon traversing a new surface, initially in the scenario where ego state estimates are converging or when transitioning between sliding and non-sliding regions of the \(\mu\)-slip curve before accurate identification can be performed.
the torque vectoring is actively exploring and adapting to changes in the road surface. 
% Although the acceleration tracking error experiences temporary tracking error, this does not impact the velocity of the overall vehicle or cause a deviation of from the original EUDC reference. That is shown in  Fig. \ref{fig:kineeudc} by also considering the ego velocity.
% This error may also be attributed to the system model's inability to capture the vehicle's behaviour at the highly non-linear limits of adhesion. 
% Hence at steady state, or operating within the non-sliding region of the tyre model, TVAL is capable of tracking the appropriate acceleration reference.
% Yet rapidly traversing between adhesion and sliding regions of the tyre model could lead to poorer performance, particularly when the estimation is poorer.
% 
% Compared to similar works \cite{Albinson2016} that operate for small slip angles, our method performs better since in this work an acceleration error of \(0.7m/s^2\) is reported at lower slip angles.
Shown in Fig. \ref{fig:tyreforce} are the tyre forces that drive the change in acceleration and show that tyre forces are highest on a dry road. 
If the estimation accuracy is lower, the vehicle can quickly generate higher acceleration error.
Yet since the tyre force maxima is better defined here, steady-state tracking is better.
% higher than is normal for TVAL, this is expected since our method is required to explore the maximum tyre forces that drives the vehicle to slide in a controlled manner.
Overall, the actions aim to minimize the exploration-exploitation objective, which in turn, improves the vehicle's ability to follow the reference speed profile.

% The tracking performance of the driver’s acceleration demand is clear since there is very little drift in the velocity and only transient spikes in acceleration (\(<0.5m/s^2\)) when the road changes or as the controller hands over to the driver. 

% At \(300s\) the road surface changes from wet to snow which requires TVAL to improve its belief. 
% However, the vehicle is operating at a high speed \(67mph\) which, when combined with the low friction road surface and poor friction estimation, leads to the system believing that the vehicle is exceeding the critical velocity \(v_{crit}\). 
% Therefore, our system is oscillating between driver only and TVAL operating modes which causes TVAL to never explore the environment fully. 
% This is seen in Fig. \ref{fig:tyreforce} where neither the front nor rear wheel tyre forces reaches the maximum.
% Despite this, TVAL only experiences an average deceleration bias error of \(-0.09m/s\).
% Oversteering is predicted at \(337s\) where the vehicle is travelling at \(78mph\). %the tyre force is also unavailable during this part of the deceleration manoeuvre where 
% At this high speed and lower friction surface (wet road), there is not enough force to achieve the drivers demand withouand achieve identification.

\begin{figure}
    \centering
    \includegraphics[scale=0.8]{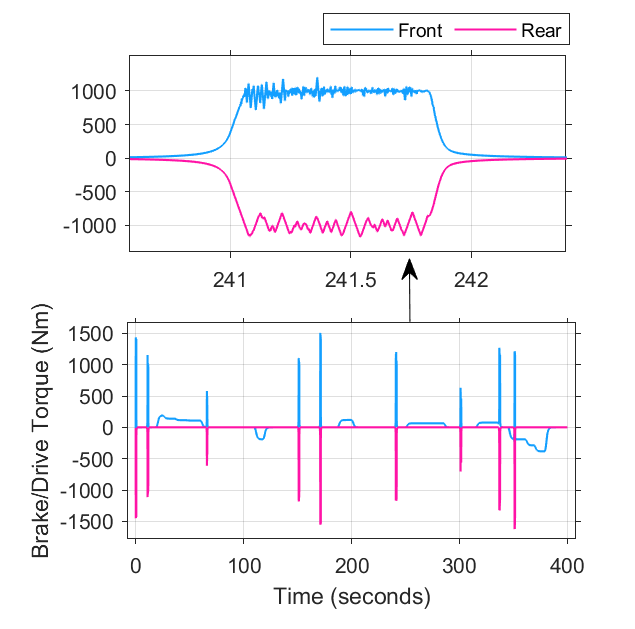}
    \caption{Front and rear wheel control input during changing road conditions.}
    \label{fig:torque}
\end{figure}

The front and rear tyre forces rise uniformly upon activation of TVAL as shown in Fig. \ref{fig:tyreforce} with the active learning weighting shown as a colormap. 
Immediately upon changing road surface, the maximum possible tyre force increases and the TVAL request begins to transition from driver to active learning. Here the tyre force slowly builds until shortly after the initial activation. 
The delay transition gradient (see (\ref{eqn:hystf}))  is set to \(\Delta P = 1\) where it can be seen that after this period, the rear tyre force reaches the maximum. 
In the example shown in Fig. \ref{fig:tyreforce}, TVAL takes only \(0.5s\) to identify the Magic formula parameters where the complete active learning cycle (driver to driver) lasts approximately 2.5s.
Noting that of this, \(2.0s\) is tunable depending on rate of transition between torque vectoring and driver inputs.
% Furthermore, as is shown in Fig. \ref{fig:tyreforce}, the tyre forces are higher on a dry road and if the estimation accuracy is poorer, the vehicle can quickly generate higher accelerations.
%
% Although the identification and stability of the observer is achieved much faster (see Fig. \ref{fig:paramd}), using the DCEE controller allows us to reduce the uncertainty and gives the system greater confidence, even as the driver may change in the input. It is this uncertainty robustness that allows us to operate at different conditions whilst maintaining a consistent augmented state belief.
This steady transition is evident in the vehicle torque inputs shown in Fig. \ref{fig:torque} where both front and rear reach a steady torque for the duration of the active learning, with a smooth transition leading up to and back to the driver.
% Although the front wheel shows a  constant torque input, the rear (which is performing the exploration-exploitation)
% The steady state oscillation while torque vectoring is due to the exploratory nature of the DCEE controller.
Notice that during this part of the manoeuvre, TVAL is continuing to perform active learning upon receiving new information.
This does not significantly affect the tyre force since, at the maximum, changes in slip (driven by the changes in the control input) do not have a significant impact.
In contrast, lower (elastic) slip ratios result in greater changes in tyre force with even small alterations in slip. 
This is clear from the slip-\(\mu\) curve depicted in Fig. \ref{fig:muslip}.

The impact of our input management mapping becomes evident at \(320s\), where the surface changes from snow to wet, prompting a brief request for TVAL (see Fig. \ref{fig:kineeudc}). 
Therefore due to the lack of tyre force availability, torque vectoring and identification of the road surface cannot be achieved.
However, by restricting the rate at which TVAL can initiate active learning, the availability constraint is violated shortly after, preventing full engagement and minimizing any potential impact from increased  torque on the vehicle.

\subsection{Friction Estimation and Tyre Force Uncertainty}
\begin{figure}
    \centering
    \includegraphics[scale=0.6]{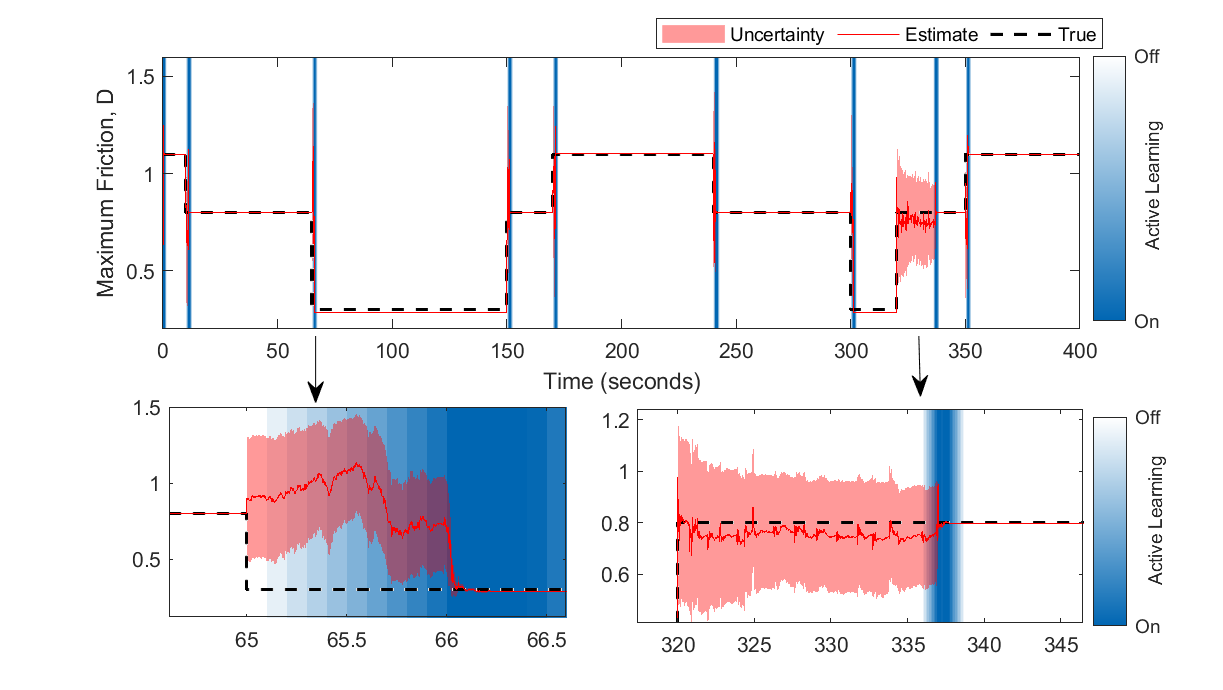}
    \caption{Estimation of the maximum road surface friction (parameter D) over changing surfaces (black dash), shown with the estimation uncertainty as shaded pink area (1 standard deviation) and active learning activation weight shown in blue. Note that the time taken to reach active learning fully on \emph{i.e.}\(w_{1,k}=1\), is tuneable where here this is set to 1 second \emph{i.e.}\(\Delta P = 1\).}
    \label{fig:paramd}
\end{figure}

% extreme weather scenario
The change of the maximum road surface friction coefficient may be considered unlikely, changing 9 times in 400s. 
However this exhibits the control systems ability to identify a range of conditions and hold its belief.
% Additionally, using 
The estimation of parameter D is shown in Fig. \ref{fig:paramd}. On low friction (snow) surfaces the maxima is less well defined that makes it the most challenging of all road surfaces selected here. 
At \(65s\) the estimation stabilizes to \(D=0.284\) giving an error of \(0.016\) (\emph{i.e.} \(5\%\)) before switching to the driver since it believes it has generated a good belief in the augmented state.
% Conversely at \(300s\), TVAL cannot reach the correct condition due to the oversteer constraint leading to an increased estimation error of \(16\%\) caused by the limited active learning effort. 
Then at \(320s\) TVAL activates for a very short period and doesn't reduce the uncertainty enough which can be noticed from the fluctuating estimation until the forward acceleration request from the driver reduces to where TVAL becomes available.
This \(320-345s\) (Fig. \ref{fig:paramd}) period shows how the estimated maximum friction and uncertainty (pink) persists until TVAL reduces the estimation error and shrinking the confidence interval before handing over to the driver in a short period of time.
%Again caused by the mostly passive effort of the filter due to the potential for oversteer.
% In all three time periods when the estimation underperforms, the estimation is always lower than than the ground truth. From a control perspective, a vehicle using this estimation would not try to overachieve the available tyre force that may be regarded as safer.

\begin{figure}
    \centering
    \includegraphics[scale=0.79]{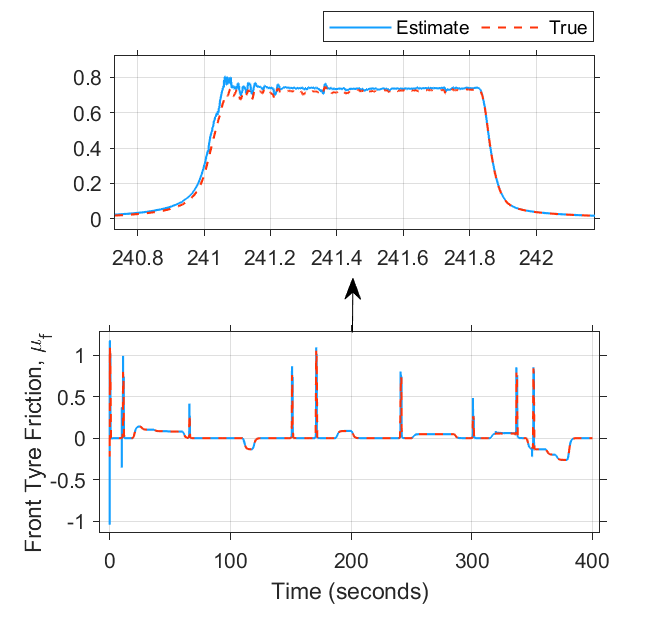}
    \caption{Rear tyre friction coefficient estimation from the Particle Filter alongside the ground truth.}
    \label{fig:frontfrict}
\end{figure}

\begin{figure}
    \centering
    \includegraphics[scale=0.79]{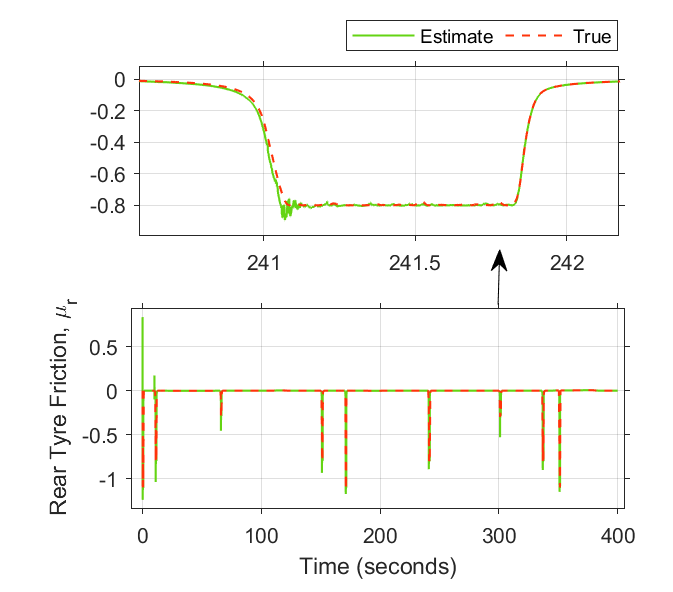}
    \caption{Front tyre friction coefficient estimation from the Particle Filter alongside the ground truth.}
    \label{fig:rearfrict}
\end{figure}

The estimated friction coefficient, shown in Fig. \ref{fig:frontfrict} and Fig. \ref{fig:rearfrict} and is estimated well throughout the scenario, even while the driver is in full control.
Performance suffers very slightly when the friction reaches the maximum on a dry road however the observer quickly converges back to the ground truth. On wet and snow surfaces, the particle filter has no issue in estimating the current friction.

\begin{figure}
    \centering
    \includegraphics[scale=0.8]{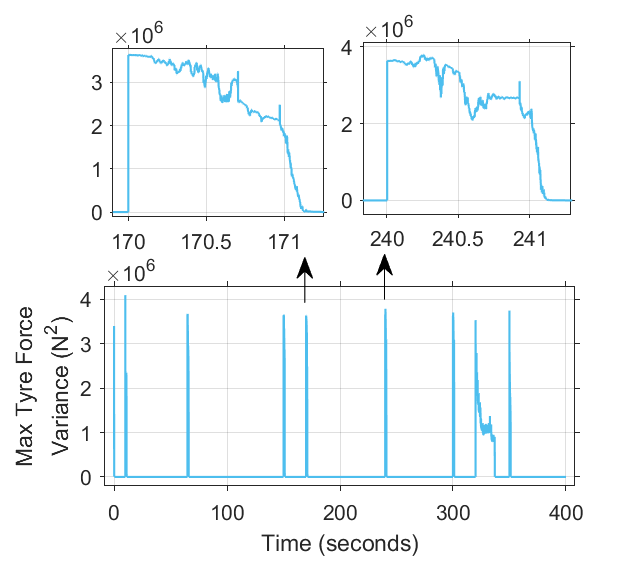}
    \caption{Maximum tyre force uncertainty found from the Particle Filter estimation.}
    \label{fig:maxF}
\end{figure}

Since this work uses a dual approach, we consider both the reference tracking and performance in reducing the uncertainty of the system. 
In Fig. \ref{fig:maxF} the estimation uncertainty in the maximum tyre force over the scenario is shown and highlights how the uncertainty decreases as the active learning (TVAL) is used (see also Fig. \ref{fig:paramd}).
In all cases where TVAL is fully active, the uncertainty decreases. One can see that during the first second (\(\Delta_p\)) the uncertainty decrease but at a slow oscilitory manner. However once TVAL becomes fully engaged (after \(1s\)) the uncertainty drastically reduces. %which is noticeable with the increased estimation performance of our friction parameters.
Notable the steady state uncertainty is reduced more than \(99.998\%\) of the original value, yielding \(6.8N\) standard deviation on the road surface staring at \(170s\).
%This certainty is significantly large when compared to the maximum ground truth tyre force of about \(6116N\).
% \begin{figure}
%     \centering
%     \includegraphics[scale=0.86]{Images/EUDC_paramd.eps}
%     \caption{Estimation performance of parameter D.}
%     \label{fig:paramd}
% \end{figure}

\subsection{Vehicle Safety}
\begin{figure}
    \centering
    \includegraphics[scale=0.8]{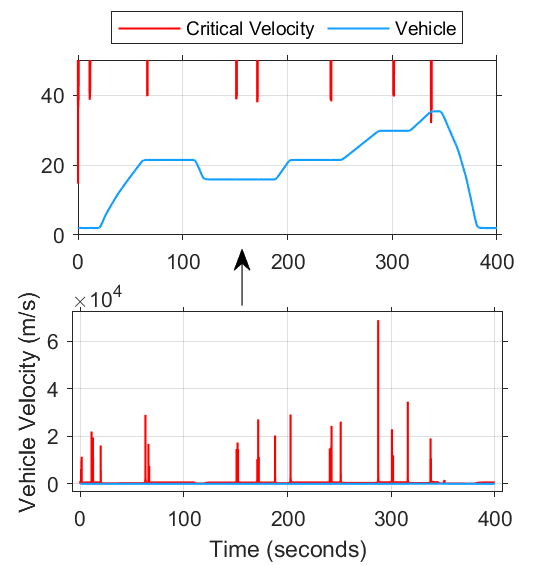}
    \caption{Critical velocity used to activate TVAL and prevent dangerous oversteer.}
    \label{fig:critvel}
\end{figure}

The estimation of the critical velocity and ego vehicle velocity are shown in Fig.~\ref{fig:critvel}.
TVAL was predicted to exceed the critical velocity at \(337s\) where the vehicle travels at high velocity on a wet road surface. 
This is to be expected given the limited availability of the tyre force.
However the vehicle is travelling at \(35.3m/s\) (\(79mph\)) that is outside normal driving for most vehicles. 
This shows the ability of our method to operate at high speeds and low speeds (\(4mph\)).
This part of the scenario is also made more challenging since the driver is decelerating here.
Therefore the simulation results show that TVAL was able to prevent the scenario that put the vehicle into a state where oversteer would have occurred if there were to be a lateral disturbance.

% Vehicle jerk is crucial for comfortable driving, particularly for prolonged autonomous journeys.
% Since the front are rear wheel control policies are dependant on the stochastic estimation of the tyre model parameters (and vehicle states), their is always a degree of error in tracking the acceleration.
% This in turn creates ego body jerk.
% Despite vehicle body jerk not being considered explicitly in the controller, the jerk is within an acceptable limit for transient operation. Shown in Fig. \ref{fig:jerk} is the vehicle body jerk over the scenario. ISO22179 suggests the negative limits on adaptive cruise control (ACC) features should be less than \(-5m/s^3\) \cite{ISO2018}.
% We argue that exceeding this temporarily does not cause intolerable discomfort. A recent survey found that brief spikes in jerk were found to have a negligible impact on the vehicle occupants \cite{DeWinkel2023}. 
% Therefore suggesting that in our case, the transient behaviour of the vehicle jerk is acceptable since the operation of TVAL is typically less than \(0.8s\).
% Furthermore, the transient tyre force behaviour was not considered here which would reduce the jerk further.
% Additionally, we can attribute the lower frequency jerk toward the end of the torque vectoring to the handover to the driver.
% Tyre wear
Tyre wear is increased when performing tyre excitation.
However, since TVAL considers the deactivation of the active learning, it is only engaged for as long as is necessary to learn the friction properties. 
Therefore the period of maximum tyre wear is typically limited to less than \(0.6s\) each time the road surface changes. 
Furthermore, high friction surfaces generate the largest wear (higher tyre forces) yet this is where the best braking performance for AEB can be achieved. 
Hence future work may consider that when high wear is experienced, active learning may be relaxed to preserve the tyres.

% \begin{figure}
%     \centering
%     \includegraphics[scale=0.8]{Images/EUDC_tyrewear.eps}
%     \caption{Tyre wear comparison between driver only and our (TVAL) method using Archard's tyre wear method. Note that during driver only mode, the vehicle is front wheel drive only compared to TVAL that is all wheel drive.}
%     \label{fig:tyrewear}
% \end{figure}
\subsection{Power and Energy Consumption}
\begin{figure}
    \centering
    \includegraphics[scale=0.85]{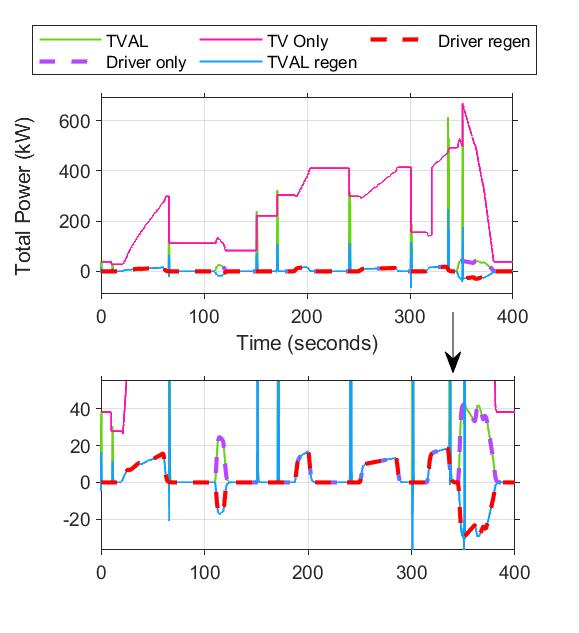}
    \caption{Vehicle power consumption shown  for our scheme (TVAL), TVAL with regenerative braking, full torque vectoring for the entire scenario \emph{i.e.} always torque vectoring (TV Only) and driver with and without torque vectoring. The full torque vectoring (pink) is computed using theoretical maximum tyre forces since full torque vectoring isn't practically achievable in this scenario which explains the transient overshoot of TVAL over the torque vectoring.}
    \label{fig:power}
\end{figure}

\begin{figure}
    \centering
    \includegraphics[scale=0.88]{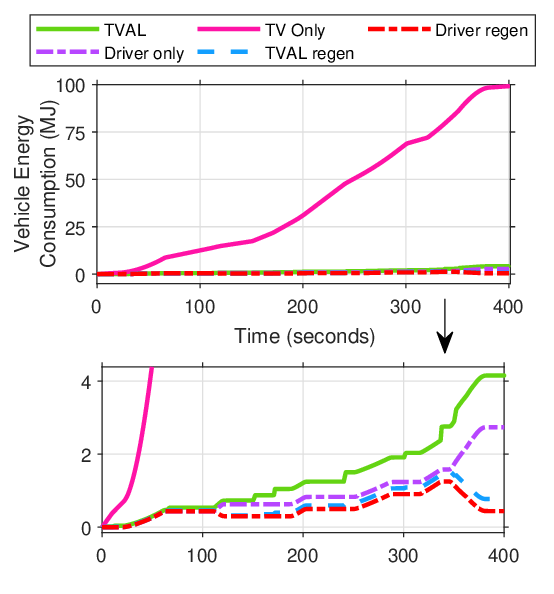}
    \caption{Cumulative power consumption shown  for our scheme (TVAL), TVAL with regenerative braking, full torque vectoring  for the entire scenario \emph{i.e.} always torque vectoring (TV Only) and driver only (no torque vectoring).}
    \label{fig:energy}
\end{figure}

% power consumption
As with any torque vectoring controller, additional energy is expended to achieve additional control tasks. The power and energy consumption is presented in Fig. \ref{fig:power} and Fig. \ref{fig:energy} respectively.
A \(96\%\) reduction in the energy consumption is made when compared to applying torque vectoring  without regulation (TVAL) \emph{i.e.} torque vectoring only. 
Additionally, TVAL requires on average \(1.5\) times the energy that normal driving requires.
One should consider that the vehicle is very unlikely to experience such challenging weather conditions, changing 9 times in 400s however this scenario is chosen to show the robustness of our method. % from 3.4
Importantly, regenerative braking, which is the recovery of energy used in braking, can be applied here (as is the case for most electric vehicles) and is also included within Fig. \ref{fig:power} and Fig. \ref{fig:energy}.
% although technological challenges exist since the battery pack cannot be discharging and charging at the same time?
Typically one may expect the efficiency of regenerative braking under heavy braking to be \(0.7\) which significantly offsets the active learning effort \cite{Doyle2017}.
This is particularly noticeable in Fig. \ref{fig:energy} where it can be seen that assuming this recovery efficiency, our method requires less energy than normal driving with non-regenerative braking.
Hence we argue that TVAL can effectively be used without causing excessive additional range anxiety to consumers, particularly given the benefit TVAL brings to active safety.
% Further shown in Fig. \ref{fig:power} is the power consumption that increases for very short periods of time when doing the exploration/exploitation action. 
A summary of the average power consumption is shown in Table \ref{tab:power}.

\begin{table}[h]
    \caption{Mean power consumption of driver and TVAL with and without regenerative braking. The theoretical constant torque vectoring (TV, not using the scheme presented here) is also shown.}
    \centering
    \begin{tabular}{|cc|}
    \hline
    Method & Average Power (\(kW\)) \\
    \hline
    TV & 247 \\
    Driver Only & 6.84 \\
    TVAL Only & 10.4\\
    TVAL with Regenerative Braking & 1.93 \\
    Driver with Regenerative Braking & 1.10 \\
    \hline
    \end{tabular}
    \label{tab:power}
\end{table}

\section{Conclusion}\label{sec:7}
TVAL is capable at identifying some of the major components required for emergency braking features such as AEB.
Knowing the current friction, peak friction coefficient and tyre model, the decision-making process for extreme braking manoeuvres is significantly improved.
Utilizing regenerative braking, we show that our work does this with less energy consumption than non-regenerative normal driving.
We look in future work to event driven TVAL including alerting the driver and delivering a holistic AEB system using this approach.
This work, although applied to an automotive use case, is applicable to the broader field of robotics for parameter identification.
% TVAL can be used where the ego vehicle reference is known and the broad behaviour of the vehicle that excites the identification of a given parameter, is also known. 
% In this case, braking or accelerating toward the friction limits.
An example for further investigation includes exploiting upcoming drive-by-wire systems, including steered rear wheels, that are being introduced into mainstream production and identifying the lateral tyre dynamics.
\end{document}